\documentclass[12pt]{article}
\usepackage{graphicx}% Include figure files

\newcommand {\be}{\begin{equation}} % start equation
\newcommand{\ee}{\end{equation}}    % end equation

\def\dds1{\frac{\partial}{\partial s_1}}

\def\nabp{\nabla_{\perp}}
\def\d{d\kern-0.8 ex\vrule height 1.3 ex depth-1.24 ex width 0.7 ex
\kern 0.15 ex}
\def\D{D\kern-1.7 ex\vrule height .87 ex depth-0.8 ex width 0.7 ex
\kern 0.95 ex}
\def\nabp{\nabla_{\perp}}

\begin{document}
\baselineskip 20 pt

\begin{center}

\Large{\bf Global convective cell formation in pair-ion   plasmas }

\end{center}

\vspace{0.7cm}

\begin{center}

 J. Vranjes

{\em Center for Plasma Astrophysics, Celestijnenlaan 200B, 3001 Leuven,
 Belgium,
and
Facult\'{e} des Sciences Appliqu\'{e}es, avenue F.D. Roosevelt 50,
 1050 Bruxelles, Belgium}

\vspace{5mm}

S. Poedts

{\em Center for Plasma Astrophysics, and Leuven Mathematical Modeling and Computational Science Center
 (LMCC), Celestijnenlaan 200B, 3001 Leuven,  Belgium,
}

\end{center}

\vspace{2cm}

{\bf Abstract:} The global electrostatic mode in pair-ion plasmas is discussed for cylindric geometry and for a radially inhomogeneous equilibrium density. In the case of a Gaussian radial density profile, exact analytical eigen solutions are found in terms of the Kummer confluent hypergeometric functions. The mode is identified as a convective cell propagating in the poloidal and axial direction, having at the same time a standing wave structure in the radial direction.

\vspace{2cm}

\noindent PACS Numbers: 52.27.Cm; 52.30.Ex; 52.35.Fp

\vspace{2cm}

\vfill

\pagebreak

A new research field and a new area of increasing scientific activity has emerged recently after a series of experiments$^{1-4}$ in which a pure pair-ion plasma has been produced, i.e.,  a plasma without electrons. The two ions, $C_{60}^{\pm}$, are produced in a simultaneous  process of impact ionization and electron attachment, and further collected by a magnetic filtering effect, i.e., by a diffusion in the radial direction (perpendicular to the magnetic field lines). Being separated from electrons, the ions are then collected in a narrow and elongated chamber (90~cm long and with a diameter of 8~cm). In Refs. 2-4 several types of modes have been reported in such plasmas, viz.\ the ion plasma wave (IPW), the ion acoustic wave (IAW), and the so-called intermediate frequency (IF) wave. The detailed measurements presented in the most recent Refs.~3 and 4, reveal that the IAW actually has two separate branches, and is accompanied by some additional backward propagating mode between these two branches. Some features of the experimentally observed modes still remain unexplained.

These experimental  works  have been followed by  numerous  studies$^{5-14}$, dealing with various aspects of linear and nonlinear waves and instabilities in pair-ion plasmas.
In addition to this, the recent successful production of a pair-hydrogen plasma,$^{4}$ that is even more attractive because of the small ion mass and the obvious consequences related to this,  indicates that the number of experiments and analytical studies in this particular field will grow even further.
The topic is  interesting  also in view of the fact that a similar pair plasma, comprising much lighter particles (electrons and positrons) in the past years has also been created under laboratory conditions.$^{15}$ The knowledge of processes in such plasmas will help us to understand the physics of some pair-plasmas in space, comprising electrons and positrons, like in the atmospheres of pulsars and in active galactic nuclei, and even in flares in the lower solar atmosphere.

So far, studies dealing with waves and instabilities$^{5-14}$ have been carried out mainly in the framework of a  local mode analysis and in Cartesian geometry. However, the experimental conditions mentioned above$^{1-4}$ are such that, in some domains of frequencies and wave lengths, the proper cylindric geometry is to be used and the effects of the plasma boundary on the modes are to be taken into account. Therefore, in the present work we shall focus on frequencies below the ion gyro-frequency, on  the effect of radial density variation (and the corresponding radial dependence of the mode amplitude), and on the effect of the boundary on the spatial structure of the mode.

We consider a cold cylindric pair-ion  plasma configuration with an externally given magnetic field in the axial direction $\vec B_0=B_0 \vec e_z$, and with a radially varying density profile $n_0(r)$. We note  that the spatial variation of the density in both the radial and axial directions has been experimentally detected,$^{1}$ with the radial density profile being nearly Gaussian. Such equilibrium  density gradients are in fact a rather common feature of various laboratory plasmas.$^{16, 17}$ Note that an axial density inhomogeneity has also been detected. Yet, in view of very different axial and radial lengths of the chamber, that inhomogeneity may be  neglected without losing much of  essential physics.

In the presence of electrostatic perturbations of the form $\sim \widehat{f}(r) \exp(-i \omega t + i \vec k \cdot \vec r + i k_z z)$, propagating at an angle with respect to the magnetic field vector,   from the momentum equations for the two species ($j=a, b$) we obtain
\be
\left(\Omega^2 +\frac{\partial^2}{\partial t^2}\right) \vec v_{j\bot 1}= \mp \frac{\Omega}{B_0} \nabla_\bot \frac{\partial \phi_1}{\partial t} + \frac{\Omega^2}{B_0} \vec e_z \times \nabla_\bot \phi_1. \label{e1}
\ee
Here, $\Omega=eB_0/m$, $m_a=m_b=m$, the $-$ sign stands for positively charged ions, and the other notation is standard. Equation~(\ref{e1}) is used in the two continuity equations which, with the help of the quasi-neutrality condition,  yield
\be
\nabp^2 \frac{\partial^2 \phi_1}{\partial t^2} + \left(\Omega^2 +\frac{\partial^2}{\partial t^2}\right) \frac{\partial^2 \phi_1}{\partial z^2}
  + \nabp \frac{\partial^2 \phi_1}{\partial t^2} \frac{\nabp n_0}{n_0}=0.
  \label{e2}
  \ee
For perturbations having a wave number component in the poloidal direction, we have $\vec k \cdot \vec r\rightarrow m \theta$ and from Eq.~(\ref{e2}) we obtain  a differential equation for  the amplitude of the perturbed potential
\be
\left[\frac{\partial^2}{\partial r^2} + \left(\frac{1}{r} + \frac{1}{n_0}\frac{dn_0}{dr}\right) \frac{\partial}{\partial r} -
\frac{m^2}{r^2} + k_z^2 \left(\frac{\Omega^2}{\omega^2}-1\right)\right] \widehat{\phi}(r)=0.
\label{e3}
\ee
Without the density gradient, the general solution of Eq.~(\ref{e3}) can be written in terms of the Bessel functions of the first kind, i.e., $\widehat{\phi}(r)= c_1 J_m[b^{1/2} r] + c_2 J_{-m}[b^{1/2} r]$, where $b=k_z^2 (\Omega^2/\omega^2 -1)$ denotes the eigenvalue. For a plasma extending in the radial direction up to  $r=R$, the proper boundary condition requires vanishing solutions at the boundary, so that $b^{1/2} R=\epsilon_l$ gives the dispersion equation. Here, $\epsilon_l$ is the $l$-th zero of the Bessel function $J_{\pm m}$.

In the presence of the density gradient, and assuming a Gaussian  density profile $n_0(r)=N_0 \exp(- \kappa^2 r^2)$, the eigenmode equation becomes
\be
\left[\frac{\partial^2}{\partial r^2} + \left(\frac{1}{r} - 2\kappa^2 r\right) \frac{\partial}{\partial r} -
\frac{m^2}{r^2} + b\right] \widehat{\phi}(r)=0.
\label{e4}
\ee
The general solution of Eq.~(\ref{e4}) is$^{18, 19}$
\be
\widehat{\phi}(r) = C_1\cdot r^{-m}\cdot {}_1F_1\left[-\frac{b}{4
\kappa_r^2}-\frac{m}{2}, 1-m, \kappa_r^2 r^2\right]
 +  C_2\cdot r^{m} \cdot
{}_1F_1\left[-\frac{b}{4 \kappa_r^2}+\frac{m}{2}, 1+m, \kappa_r^2
 r^2\right].
 \label{e5} \ee
Here, ${}_1F_1$ is the Kummer confluent hypergeometric function, and the $C_{1, 2}$ correspond to the integration constants.

The eigenvalue $b$ yields the eigenfrequency
\be
\omega^2=\Omega^2\, \frac{k_z^2}{b + k_z^2}, \label{e6}
\ee
which describes the obliquely propagating (in the $\theta$ and $z$ directions) electrostatic convective cells, i.e., modes twisted around the magnetic field vector. Examples of similar convective cells in the domain of the ion gyro-frequency, and the lower-hybrid frequency, are known from the literature.$^{20}$

Remarkable properties of the pair-ion plasma can be seen by comparing   Eqs.~(\ref{e5} and \ref{e6}) with the corresponding equations in  electron-ion plasmas,$^{18}$, later describing the drift wave.
In the present case, the poloidal wave number $m$ directly determines only the radial eigenfunctions. The same holds for the equilibrium density profile parameter $\kappa$. The dispersive properties of the mode are directly determined only by the eigenvalues $b$.  However, in the electron-ion plasma$^{18}$ the dispersion equation contains all the parameters, viz.\ the poloidal wave number $m$, the density parameter $\kappa$, and the eigenvalue $b$. The difference between the two cases appears due to the pair property of the plasma in the present case, which includes the cancelation of terms with the $\vec E\times \vec B$-drift, i.e., the terms  $\vec v_E \nabla n_0$,  in two combined  continuity equations.

Well behaving  solutions are finite in the plasma column and should  not have a finite radial velocity at $r=R$ and at $r=0$. Hence, $\partial \widehat{\phi}_1(r=0, 1)/\partial \theta=0$, and this is  satisfied with a potential profile vanishing at the axis and the boundary. Therefore, we set $C_1=0$, and the appropriate profile will be found for certain values of the parameters $\kappa$ and  $b$.

Observe that the  function $ {}_1F_1[d_1, d_2, r]$ is a polynomial with a finite number of terms if $d_1\leq 0$, and if in the same time $d_2>0$ or $d_2<d_1$. This implies oscillatory (standing wave) solutions in the radial direction. In the present case, we have $d_2=1+m>0$, and oscillatory solutions $d_1\leq 0$ exist if $b\geq 2 m \kappa^2$. For example, $ {}_1F_1[-1, 2, r^2]=1-r^2/2$, \, $ {}_1F_1[-2, 2, r^2]=1-r^2 + r^4/6$,\, $ {}_1F_1[-3, 2, r^2]=1-3r^2/2+ r^4/2 - r^6/24$.

The eigen-values $b$ satisfying the boundary conditions (e.g., a vanishing potential at $r=R$) can be found numerically. There exists a multiple choice of eigenvalues $b$, in terms of $\kappa$, as shown in our recent Refs.~21, 22. The values of $\kappa$ may be taken from  the interval  0.1 to  1.5. It is easily seen that for the assumed Gaussian density $n(\kappa, r)$ normalized to $N_0$, where  $r$ is normalized to $R$, this yields $n(0.1,1)=0.99$ and $n(1.5,1)=0.1$, respectively. For this range of $\kappa$, and taking as example  the poloidal mode number $m=3$, the first branch of the eigenvalues $b$ in terms of $\kappa$ is presented in Fig.~1. These pairs of  $b, \kappa$ give the first eigenfunction $ r^{3} \cdot {}_1F_1$, i.e., the profile   with  the first node  of the radially oscillating standing wave solution ar $r=R$.
It can be shown that for $m=3$ and for $\kappa=0.1$, the second branch starts with $b=95.3$. This branch gives the second eigenfunction,  the one with the  second node of   $ r^{3} \cdot {}_1F_1$ at $r=R$.  The third branch begins with  $b=169.2$, etc.

The three first eigenfunctions (standing wave solutions in the radial direction)  are presented in Fig.~2 for $\kappa=0.3$, and $b= 40.53,\, 95.18, \, 169.2$, respectively. A full three dimensional plot of the modes gives twisted solutions of the form $\cos(m \theta + k_z z - \omega t)\cdot r^m \cdot {}_1F_1$, i.e., electrostatic waves traveling in the poloidal and axial directions, having at the same time a standing wave structure in the radial direction.

The dispersion equation, Eq.~ (\ref{e6}), shows that the frequency of the mode remains below the ion gyro-frequency. Each discrete eigenvalue $b$ gives a separate dispersion line, the first three lines being presented in Fig.~3. Hence, for the given density profile  with $\kappa=0.3$, any of these branches of oscillations can take place.

To conclude, in this Brief Communication exact analytical solutions are presented, that should be expected in the low frequency range below the ion gyro-frequency in the recently produced radially inhomogeneous pair-ion plasma. Contrary to the similar case in electron-ion plasma, where in this frequency range the drift mode is found,$^{18, 19}$ in the present case, because of the pair properties of the plasma components, these modes are convective cells, and the mode frequency is not directly determined by the density gradient. Global modes determine global properties of a plasma, so that after the  identification of the modes described here, their eventual observation should be used in the diagnostics of pair-ion plasmas.

\vspace{.5cm}

\paragraph{Acknowledgements:}
The  results presented here  are  obtained in the framework of the
projects G.0304.07 (FWO-Vlaanderen), C~90205 (Prodex),  GOA/2004/01
(K.U.Leuven),  and the Interuniversity Attraction Poles Programme -
 Belgian State - Belgian Science Policy.

\vfill

\pagebreak

\noindent{\bf Figure captions:}

\begin{description}
\item{Fig.~1.} Locus of pairs $b, \kappa$ for the first eigenfunction,  satisfying the condition $r^{3}\cdot {}_1F_1=0$ at $r=R$.
\item{Fig.~2.} The three lowest radial eigenfunctions $r^{3}\cdot {}_1F_1$, in arbitrary units, for $\kappa=0.3$.
\item{Fig.~3.} Normalized frequency (\ref{e6}) for the three first eigenfunctions from Fig.~2,  for the given density profile with $\kappa=0.3$.

\end{description}

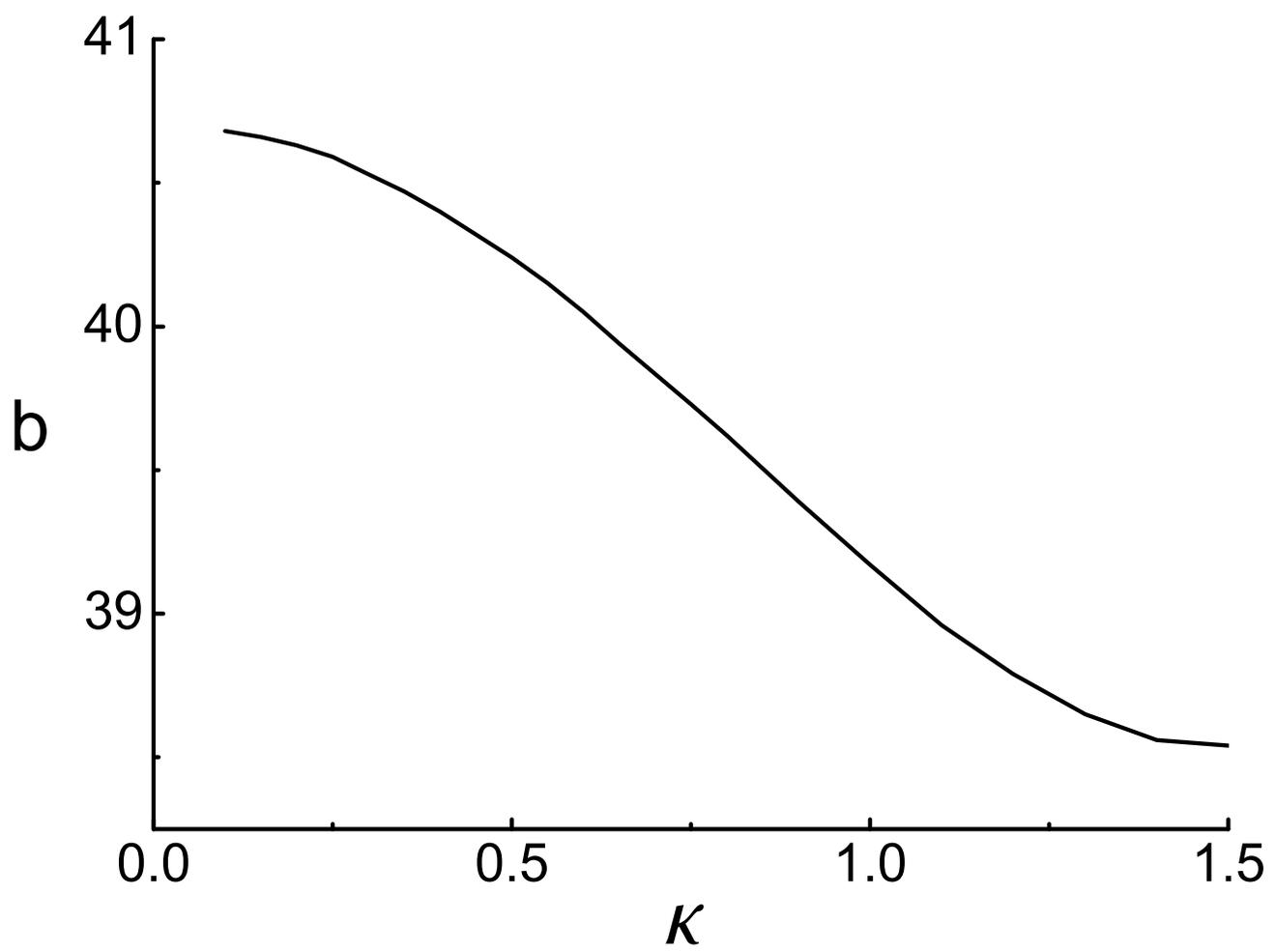
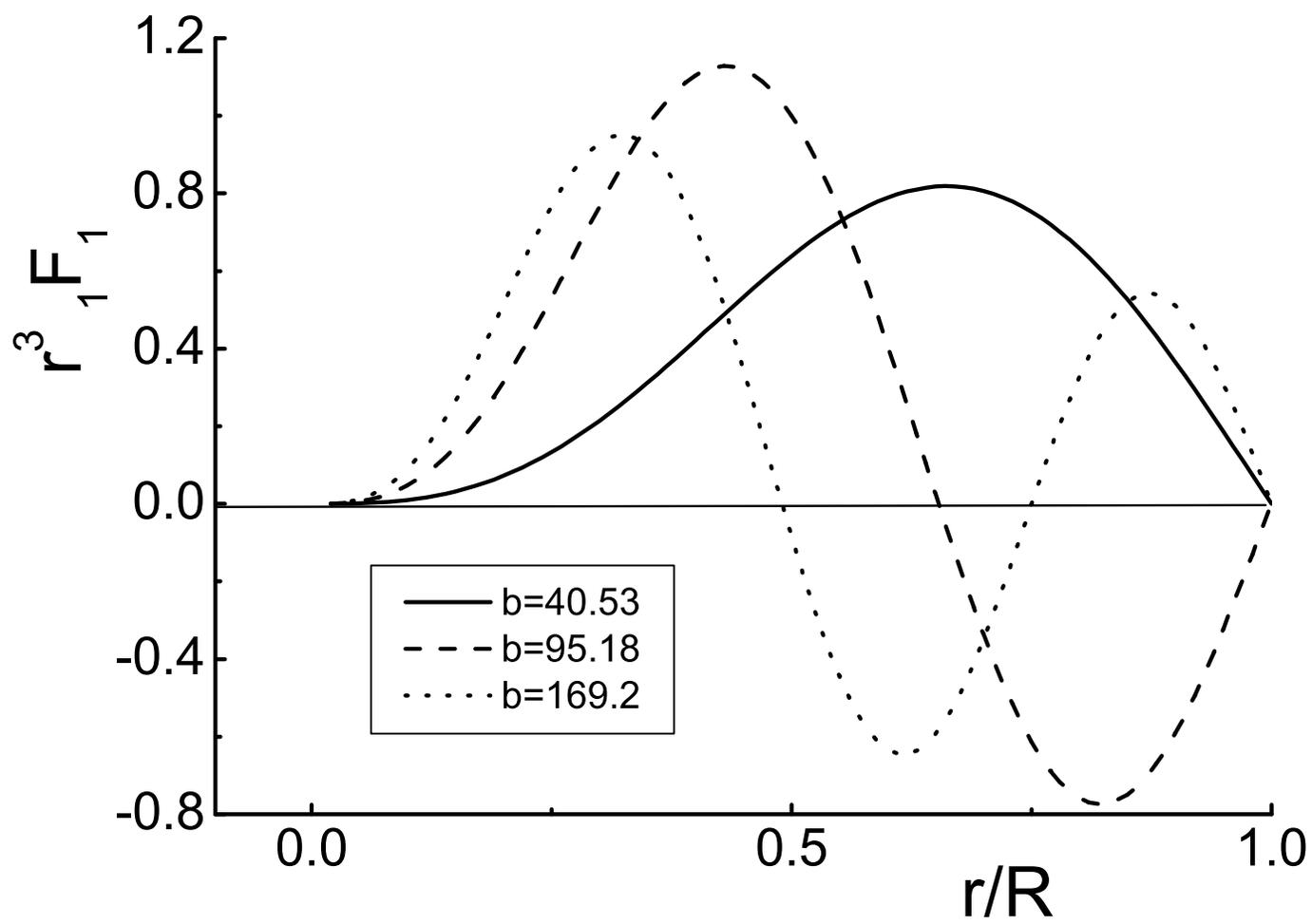
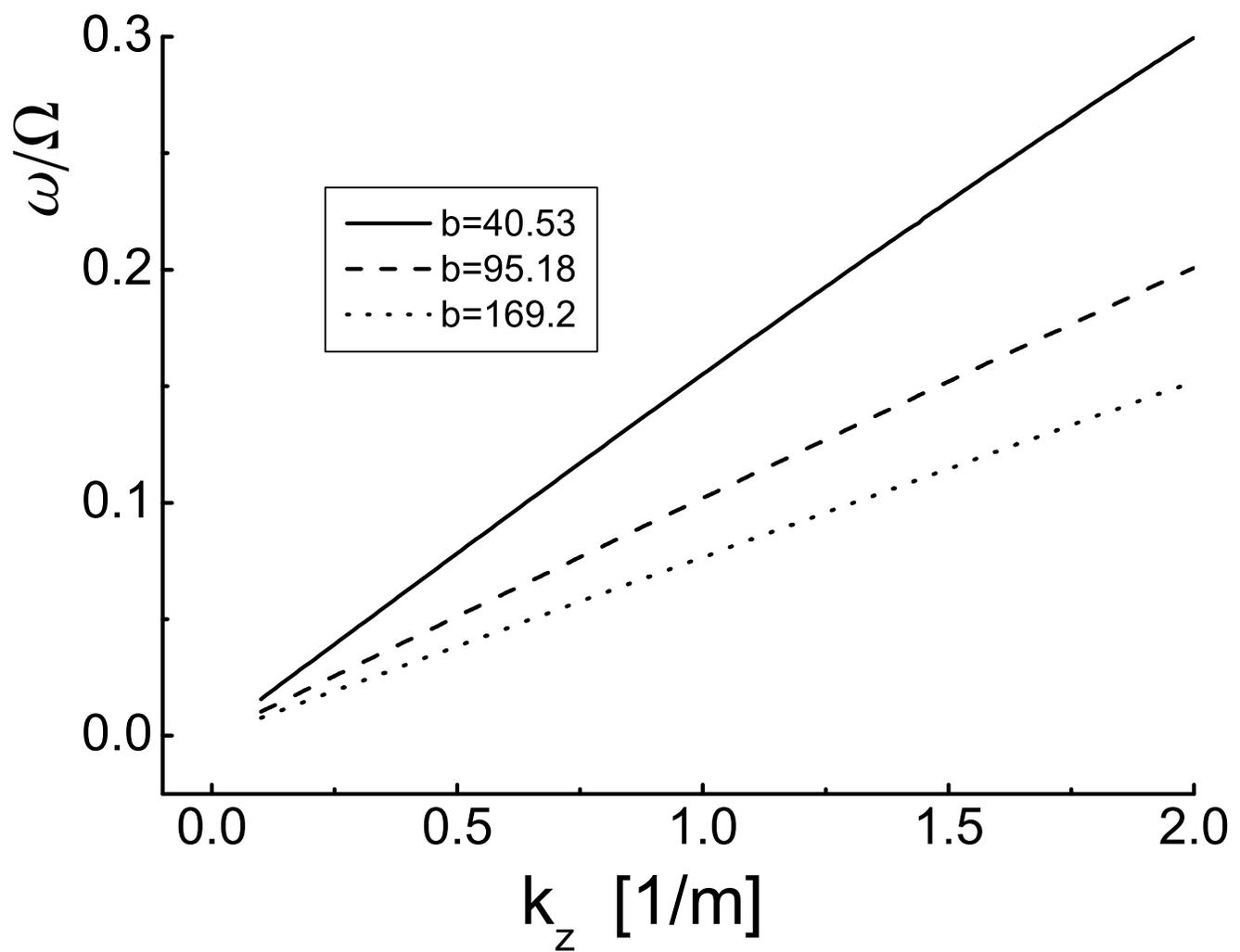
\end{document}